\documentclass[12pt,english,floatfix,showkeys,superscriptaddress,aps,prd,preprint]{revtex4}
\usepackage[latin1]{inputenc}
\usepackage[T1]{fontenc}
\usepackage{lmodern}
\usepackage{amsmath}
\usepackage{amssymb}
\usepackage{graphicx}
\usepackage{float}
\usepackage{esint}
\usepackage{dcolumn}
\usepackage{babel}
\usepackage{csquotes}
\usepackage{color}
\usepackage{xcolor}
\usepackage{caption}
\usepackage{subcaption}
\usepackage{slashed}
\usepackage{simplewick}
\usepackage{tikz-feynman}
\usepackage[]{tikz-feynman}
\usepackage{amsmath,latexsym}
\usepackage{ragged2e}

\usepackage{cancel}

\usepackage{orcidlink}
\usepackage{comment}

\usepackage{hyperref}
\hypersetup{
    colorlinks,
    citecolor=blue,
    filecolor=green,
    linkcolor=purple,
    urlcolor=red,
}

\usepackage{slashed}

\usepackage{hyperref}
\hypersetup{colorlinks,breaklinks,
			citecolor=[rgb]{0,0.0,1.0},
            urlcolor=[rgb]{0.0,0.0,1.0},
            linkcolor=[rgb]{0,0.5,0.9}}

\begin{document}

\title{Fermion-fermion scattering in a Rarita-Schwinger model with Yukawa-like interaction}

\author{M. C. Ara\'{u}jo\orcidlink{0009-0009-8703-6092}}
\email{michelangelo.araujo@ufca.edu.br}
\affiliation{Universidade Federal do Cariri, Av. Tenente Raimundo Rocha, \\ 
Cidade Universit\'{a}ria, Juazeiro do Norte, Cear\'{a}, 63048-080, Brazil}
\author{J. G. Lima\orcidlink{0000-0002-2443-0043}}
\email{junior.lima@uaf.ufcg.edu.br}
\affiliation{Departamento de F\'{\i}sica, Universidade Federal de Campina Grande,\\
Caixa Postal 10071, 58429-900, Campina Grande, Para\'{\i}ba, Brazil}
\author{J. Furtado\orcidlink{0000-0002-1273-519X}}
\email{job.furtado@ufca.edu.br}
\affiliation{Universidade Federal do Cariri, Av. Tenente Raimundo Rocha, \\ 
Cidade Universit\'{a}ria, Juazeiro do Norte, Cear\'{a}, 63048-080, Brazil}
\author{T. Mariz\orcidlink{0000-0002-3610-2557}}
\email{tmariz@fis.ufal.br}
\affiliation{Instituto de F\'{i}sica, Universidade Federal de Alagoas, Macei\'{o}, Alagoas, 57072-900, Brazil}
\date{\today}

\begin{abstract}
In this work, we investigate the scattering of spin-$3/2$ fermionic particles mediated by a Yukawa-like coupling in the context of the massive Rarita-Schwinger model. The interaction is introduced by replacing $m \to m_{\psi} + g\phi$ in the free spin-$3/2$ Lagrangian. The analysis is performed at both zero and finite temperatures. In the latter case, thermal effects are incorporated using the Thermofield Dynamics (TFD) formalism. In both regimes, we obtain the differential and total cross sections and examine their behavior in the short-range ($m_{\phi} \neq 0$) and long-range ($m_{\phi} = 0$) limits, in order to analyze the influence of zero- and finite-temperature effects.
\end{abstract}


\maketitle


\section{Introduction}\label{intro}

Studies of higher-spin field theories, especially those involving fields with spin $\geq 3/2$, have been extensively explored in the literature \cite{Bengtsson:2020,Bonora:2016otz,Bonora:2017ykb,Rahman:2012thy}. Essentially, this type of study is relevant for self-consistent models of grand unification theories \cite{Adler:2014pga}, for improving the understanding of the ultraviolet (UV) behavior of gravitational interactions \cite{Campoleoni:2024ced}, and for naturally explaining the emergence of these theories in the AdS/CFT correspondence \cite{Giombi:2016ejx}, among other scenarios. Although free propagation of arbitrary-spin fields is generally possible, the nature of their interactions strongly depends on the spin \cite{Singh:1974qz,Singh:1974rc,Fronsdal:1978rb,Fang:1978wz}. Indeed, the introduction of interactions often leads to consistency challenges, such as the loss of the correct physical degrees of freedom for higher-spin particles.

The Rarita-Schwinger (RS) model, which describes spin-$3/2$ fields \cite{Rarita:1941mf}, has attracted considerable phenomenological interest in various contexts. This model plays a central role in supergravity (SUGRA) \cite{Freedman:1976py,Das:1976ct,Gates:1983nr}, for the description of the gravitino (for a recent review, see \cite{Nath:2026yma}), and in the modeling of hadronic resonances \cite{deJong:1992wm,Pascalutsa:1999zz,Bernard:2003xf}. It is also employed, e.g., in scattering processes involving spin-$3/2$ particles \cite{Delgado-Acosta:2009ulg,Antoniadis:2022jjy}, in studies of structures that violate Lorentz symmetry \cite{Gomes:2022btc,Gomes:2023qkj,Gomes:2024qya}, and within the framework of very special relativity \cite{Araujo:2025tgy}. Despite its substantial physical contributions and strong phenomenological impact, the interacting RS model imposes consistency constraints that continue to be investigated in the literature \cite{Moldauer:1956zz,Johnson:1960vt,Velo:1969txo,Velo:1969bt,Aurilia:1969bg,Nath:1971wp,Adler:2015yha,Adler:2015zha}. 

Among the difficulties associated with the RS model, the introduction of interactions proves to be restricted. In particular, in the presence of certain types of coupling, e.g., external fields, quantization may require structures incompatible with the usual hypothesis of a positive-definite metric, and even at the classical level, non-causal propagation modes may arise \cite{Velo:1969bt,Johnson:1960vt}. In this context, the search for a consistent coupling prescription becomes relevant. One possibility that has been discussed for consistently introducing a scalar coupling is to generate the interaction from the mass term by promoting the mass term to a field-dependent quantity, i.e., $m \to m + g\phi$ \cite{Hagen:1974sn}.

From a theoretical perspective, the Yukawa theory \cite{Yukawa:1935xg} provides a scenario in which it is possible to investigate interactions mediated by a scalar boson. This mechanism is responsible for the scattering process that gives rise to the Yukawa potential \cite{peskin,ryder}. This theory was highly successful and was the first model of the nuclear force leading to the development of the theory of mesons \cite{Nambu}. As a result, several works in the literature address the Yukawa interaction in various settings \cite{Yukawa1,Yukawa2,Yukawa3,Yukawa4}. In this context, it becomes relevant to explore the interactions of Yukawa theory in the RS model, i.e., the Yukawa-like interactions involving spin-$3/2$ fields.

Studies of scattering processes under the influence of thermal effects have aroused significant interest, both in real-time and imaginary-time formulations \cite{Matsubara:1955ws,Takahashi:1996zn,Umezawa:1982nv,khannatfd,Das:1987yb,Dolan:1973qd,Das:2010fp,Das:2011yj}. In real time, the thermofield dynamics (TFD) formalism proves convenient for tree-level scattering investigations, encompassing both spinorial \cite{Cabral:2024tqa, Santos:2021hed, Souza:2021uuf} and scalar \cite{Araujo:2024hnu,Araujo:2022qke} processes. Furthermore, thermal effects in Lorentz violation extensions of non-abelian gauge sectors have also been analyzed using this formalism \cite{Santos:2022zfz}. The TFD formalism is based on two fundamental concepts: the duplication of the Hilbert space and the Bogoliubov transformation \cite{Takahashi:1996zn,Umezawa:1982nv}. A consequence of this is that for the propagator of any theory, it is divided into two parts: the standard propagator at zero temperature and a thermal component.

The analysis of Yukawa interactions with Lorentz violation via TFD was recently explored in \cite{Cabral:2024fhg}. However, scattering processes involving spin-$3/2$ particles under the influence of thermal effects have been little explored in the literature. In a previous work \cite{Araujo:2024bug}, the authors explored the impacts of finite temperature effects on Bhabha-like scattering in the RS model using TFD formalism.

Our aim in the present work is to study Yukawa-like scattering in a massive RS model at zero and finite temperature. The Yukawa scattering is investigated at tree level, and the effects of finite temperature are addressed using the TFD formalism. We focus on standard fermion-fermion scattering that includes a Yukawa-like interaction through the replacement $m \to m_\psi+g\phi$ in the spin-$3/2$ Lagrangian, where the corresponding cross sections are calculated.

The present article is structured as follows. In Section \ref{RS}, we discuss the Lagrangian of the RS field coupled to Yukawa theory, including its general form, plane-wave expansion, Feynman rules, and the calculation of the Yukawa-like differential cross section at zero temperature. The TFD formalism for this model is presented in Section \ref{TFD}, where the differential cross section for the Yukawa-like interaction at finite temperature is calculated using this formalism. Finally, in Section \ref{conclusion}, we summarize the main results of the paper. Throughout the article, we use natural units and assume $\eta^{\mu\nu} = \mathrm{diag}(1,-1,-1,-1)$ as the Minkowski metric.

\section{Analysis at zero temperature}\label{RS}

\subsection{The model}

The Lagrangian density describing free spin-$3/2$ fermions, in its most general form, is given by 
\begin{eqnarray}\label{freedensitylagferm}
\mathcal{L} = \Bar{\psi}^{\mu}\Lambda_{\mu \nu}\psi^{\nu},
\end{eqnarray}
where 
\begin{eqnarray}\label{generalcaseL}
 \Lambda^{\mu\nu}&=&(i\slashed{\partial}-m)\eta^{\mu\nu}
+iA(\gamma^{\mu}\partial^{\nu}+\gamma^{\nu}\partial^{\mu})+\frac{1}{2}(3A^2+2A+1)\gamma^{\mu}i\slashed{\partial}\gamma^{\nu} \nonumber\\
&+& m(3A^2+3A+1)\gamma^{\mu}\gamma^{\nu},
\end{eqnarray}
with $A$ being an arbitrary real parameter. In its original formulation, first proposed in \cite{Rarita:1941mf}, the value assumed for $A$ was $-1/3$. Here, however, we consider the specific case where $A = -1$, since it corresponds to the definition most commonly adopted in the literature nowadays. With this choice, the free Lagrangian density in Eq.~\eqref{freedensitylagferm} takes the form 
\begin{eqnarray}\label{freedensitylagfermamenos1}
\mathcal{L} = \Bar{\psi}_{\mu}\, \frac{i}{2}\{\sigma^{\mu\nu}, (i\slashed{\partial} -m) \}\, \psi_{\nu},
\end{eqnarray} where $\sigma^{\mu\nu}=\frac{i}{2}[\gamma^{\mu},\gamma^{\nu}]$ is the Dirac sigma matrix. Now, in order to allow for the coupling between the fermionic field and a scalar field, we perform the replacement $m \rightarrow m_{\psi}+g\phi$, with $m_{\psi}$ denoting the fermion mass and $g$ a dimensionless coupling parameter \cite{Hagen:1974sn}. After these considerations, we can write
\begin{eqnarray}\label{densityofinterest}
\mathcal{L} = \Bar{\psi}_{\mu}\, \frac{i}{2}\{\sigma^{\mu\nu}, (i\slashed{\partial} -m_{\psi}-g\phi) \}\, \psi_{\nu} +\frac{1}{2}(\partial_{\mu}\phi \partial^{\mu}\phi - m_{\phi}^2\phi^2),
\end{eqnarray}
where the contribution of the free scalar field with mass $m_{\phi}$ has also been included. 

Taking as a starting point the Lagrangian density in Eq.~\eqref{densityofinterest}, we can straightforwardly derive the Feynman rules. The propagators for the fermionic and scalar fields, for instance, are given respectively by
\begin{eqnarray}
    S^{\mu \nu}(q) = i \frac{\slashed{q}+m_{\psi}}{q^2-m_{\psi}^2}\Bigg( \eta^{\mu \nu} -\frac{1}{3}\gamma^{\mu}\gamma^{\nu}-\frac{1}{3}\frac{\gamma^{\mu}q^{\nu}-\gamma^{\nu}q^{\mu}}{m_{\psi}}-\frac{2}{3}\frac{q^{\mu}q^{\nu}}{m_{\psi}^2} \Bigg)
\end{eqnarray}
and 
\begin{eqnarray}\label{propescalarmasphi}
    D(q)=\frac{i}{q^2-m_{\phi}^2}.
\end{eqnarray}
At this point, a remark is in order. Note that when $m_{\psi} = 0$, the operator $\Lambda ^{\mu \nu}$ becomes transverse and the theory exhibits a gauge symmetry associated with the field $\psi^{\mu}$, so that a gauge-fixing term is required in order to properly define the fermionic propagator. When $m \neq 0$, no gauge-fixing term is necessary, since the gauge symmetry is broken. Although this distinction may lead to completely different scenarios, it will not affect the scattering process we are interested in investigating. This is because, as we shall see, the interactions considered here are mediated by scalar bosons, which ensures that our results remain valid in both regimes. The interaction vertex, in turn, is given by 
\begin{eqnarray}\label{verticemunuetagammagamma}
    V^{\mu \nu}= -ig(\eta^{\mu \nu}-\gamma^{\mu}\gamma^{\nu}),
\end{eqnarray}
which differs from that of the usual Yukawa theory due to the presence of the metric tensor $\eta^{\mu \nu}$ and an additional term involving the product of Dirac gamma matrices.

\subsection{Yukawa-like scattering}

\begin{figure}
    \centering
    \begin{tikzpicture}[baseline=(b3.base)]
 \begin{feynman}
    \vertex (a1);
    \vertex [right=1.3 of a1] (a2);
    \vertex [right=1.3 of a2] (a3);
    \vertex [right=1.3 of a3] (a4);
    \vertex [below=1.3 of a1] (b1);
    \vertex [right=1.3 of b1] (b2);
    \vertex [right=1.3 of b2] (b3);
    \vertex [right=1.3 of b3] (b4);
    \vertex [below=1.3 of b1] (c1);
    \vertex [right=1.3 of c1] (c2);
    \vertex [right=1.3 of c2] (c3);
    \vertex [right=1.3 of c3] (c4);

    \diagram* {
      (a1) -- [anti fermion, edge label'=\(k_1\)] (b2),
      (b2) -- [scalar, edge label'=\(q\)] (b3), 
      (b3) -- [fermion, edge label'=\(k_2\)] (a4),
      (c1) -- [fermion, edge label=\(p_1\)] (b2),
      (b3) -- [anti fermion, edge label=\(p_2\)] (c4),
    };
  \end{feynman}
 \end{tikzpicture}
 \hspace{1cm}
 \begin{tikzpicture}[baseline=(b3.base)]
 \begin{feynman}
    \vertex (a1);
    \vertex [right=1.3 of a1] (a2);
    \vertex [right=1.3 of a2] (a3);
    \vertex [right=1.3 of a3] (a4);
    \vertex [below=1.3 of a1] (b1);
    \vertex [right=1.3 of b1] (b2);
    \vertex [right=1.3 of b2] (b3);
    \vertex [right=1.3 of b3] (b4);
    \vertex [below=1.3 of b1] (c1);
    \vertex [right=1.3 of c1] (c2);
    \vertex [right=1.3 of c2] (c3);
    \vertex [right=1.3 of c3] (c4);

    \diagram* {
      (a1) -- [anti fermion, edge label'=\(k_1\)] (b3),
      (b2) -- [scalar, edge label'=\(q'\)] (b3), 
      (b2) -- [fermion, edge label'=\(k_2\)] (a4),
      (c1) -- [fermion, edge label=\(p_1\)] (b2),
      (b3) -- [anti fermion, edge label=\(p_2\)] (c4),
    };
    \draw [white, line width=0.9pt]
  ($(b2)!0.2!(a4)$) -- ($(b2)!0.30!(a4)$);
  \draw [black, line width=0.42pt]
  ($(b3)!0.23!(a1)$) -- ($(b3)!0.30!(a1)$);
  \end{feynman}
 \end{tikzpicture}
    \caption{Feynman diagrams for fermion-fermion scattering in the Yukawa theory.}
   \label{diagrelevantes}
\end{figure}
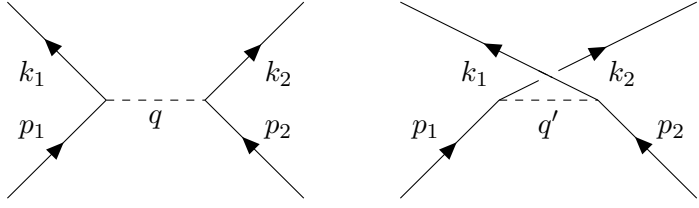

In this section, we apply the Feynman rules discussed previously to the two-particle scattering process characterized by
\begin{eqnarray}\label{fermionfermionsctteringyukawa}
\text{fermion}\, (p_1) + \text{fermion}\, (p_2) \rightarrow \text{fermion}\, (k_1) + \text{fermion}\, (k_2).
\end{eqnarray}
At tree level, the two Feynman diagrams shown in Fig.~\ref{diagrelevantes} contribute, with the first corresponding to the $t$-channel interaction and the second to the $u$-channel. As can be seen, this process is mediated by a scalar field $\phi$, whose propagator is given in Eq.~\eqref{propescalarmasphi}. 

Since all fermions involved in the process have identical masses, the differential cross section, when evaluated in the center-of-mass frame, takes the form 
\begin{eqnarray}\label{difseczerotemspin}
      \frac{\mathrm{d} \sigma }{\mathrm{d}\Omega} = \frac{1}{64\, \pi^2 \, E_{cm}^2}\, \frac{1}{16}\sum_{spin}|\mathcal{M}|^2,
\end{eqnarray} 
where the scattering matrix $\mathcal{M}$, taking into account the two interaction channels discussed above and constructed in the usual way using the Feynman rules, is given by
\begin{eqnarray}
    \mathcal{M} = \mathcal{M}_t+\mathcal{M}_u,
\end{eqnarray}
with
\begin{eqnarray}
    \mathcal{M}_t = \frac{1}{q^2-m_{\phi}^2}\mathbf{\bar{u}}_{\mu}^{r_1}(k_1)\, V^{\mu \nu}\, \mathbf{u}_{\nu}^{s_1}(p_1)\, \mathbf{\bar{u}}_{\mu'}^{r_2}(k_2)\, V^{\mu' \nu'}\, \mathbf{u}_{\nu'}^{s_2}(p_2)
\end{eqnarray}
and
\begin{eqnarray}
    \mathcal{M}_u = -\frac{1}{q'^2-m_{\phi}^2}\mathbf{\bar{u}}_{\mu}^{r_2}(k_2)\, V^{\mu \nu}\, \mathbf{u}_{\nu}^{s_1}(p_1)\, \mathbf{\bar{u}}_{\mu'}^{r_1}(k_1)\, V^{\mu' \nu'}\, \mathbf{u}_{\nu'}^{s_2}(p_2).
\end{eqnarray}
It is worth emphasizing that $\mathbf{u}_{\mu}$ and $\mathbf{\bar{u}}_{\mu}$ follow directly from the plane-wave expansion of the fermionic field, conventionally written as 
\begin{eqnarray}\label{campopsimuxexpan}
       \psi_{\mu}(x) &=& \int \frac{d^3p}{(2\pi)^3}\frac{1}{\sqrt{2E(\mathbf{p})}}\sum_s \left[ a^s(\mathbf{p})\mathbf{u}_{\mu}^s(p)e^{-ip\cdot x}+b^{s\, \dagger}(\mathbf{p})\mathbf{v}_{\mu}^s(p)e^{ip\cdot x} \right],
\end{eqnarray}
where $\mathbf{u}_{\mu}$ and $\mathbf{v}_{\mu}$ denote the Rarita-Schwinger spinors, while $a$ and $b^{\dagger}$ correspond to the annihilation and creation operators for spin-$3/2$ fermions and antifermions, respectively.

Now, since in the center-of-mass reference frame we have
\begin{eqnarray}\label{setcentermassframe33}
p_1=(E,|\mathbf{p}|\, \hat{\mathbf{z}}); &\hspace{1cm}& k_1=(E,\mathbf{k}); \nonumber\\
p_2=(E,-|\mathbf{p}|\, \hat{\mathbf{z}}); &\hspace{1cm}&  k_2=(E,-\mathbf{k}); 
\end{eqnarray} with
\begin{eqnarray}
 \mathbf{k} = |\mathbf{k}|(\sin \theta \, \cos \phi \, \hat{\mathbf{x}}+\sin \theta \, \sin \phi \, \hat{\mathbf{y}}+ \cos \theta \, \hat{\mathbf{z}}),
\end{eqnarray}
and
\begin{eqnarray}
   |\mathbf{p}|=|\mathbf{k}|=\sqrt{E^2-m_{\psi}^2}, 
\end{eqnarray}
the differential cross section in Eq.~\eqref{difseczerotemspin} can be written as
\begin{eqnarray}\label{secdifzerotempfinal1}
    \frac{\mathrm{d} \sigma}{\mathrm{d} \Omega}  &=&  \frac{g^4}{41472\, \pi ^2\, E^2\, m_{\psi }^8}\, \Pi_0(m,E,\theta),
\end{eqnarray}
where
\begin{eqnarray}\label{deffuncpizerometheta}
  \Pi_0(m,E,\theta)  &=& \frac{-1}{[m_{\phi }^2-16 \sin^4\left(\frac{\theta}{2}\right) (E^2-m_{\psi }^2)^2]^2 [m_{\phi }^2-16
   \cos^4\left(\frac{\theta}{2}\right) (E^2-m_{\psi }^2)^2]^2}\nonumber\\
   &\times& [(E^2-m_\psi^2)^4 (-256 E^{12} \sin ^8(\theta) (\cos (2 \theta )+7)\nonumber\\
   &+& 512 E^{10} \sin^8(\theta)
   (3 \cos (2 \theta )+29) m_{\psi }^2\nonumber\\
   &+& 64 E^8 \sin ^6(\theta ) (590 \cos (2 \theta )+15 \cos (4 \theta )-1277) m_{\psi }^4\nonumber\\
   &+& 64
   E^6 \sin ^4(\theta ) (-1211 \cos (2 \theta )+380 \cos (4 \theta )+5 \cos (6 \theta )+1114) m_{\psi }^6\nonumber\\
   &-& 16 E^4 \sin
   ^4(\theta ) (9103 \cos (2 \theta )+1970 \cos (4 \theta )+15 \cos (6 \theta )+15120) m_{\psi }^8\nonumber\\
   &-& 8 E^2 \sin ^2(\theta ) (48300 \cos
   (2 \theta )+8622 \cos (4 \theta )+604 \cos (6 \theta )\nonumber\\
   &+& 3 \cos (8 \theta )+39239) m_{\psi }^{10}-(319348 \cos (2 \theta )+80908 \cos (4
   \theta )\nonumber\\
   &+& 7915 \cos (6 \theta )+288 \cos (8 \theta )+\cos (10 \theta )+255092) m_{\psi }^{12})\nonumber\\
   &+& m_{\phi }^4 (-934
   E^{12}+900 E^{10} m_{\psi }^2-4360 E^8 m_{\psi }^4+5736 E^6 m_{\psi }^6-3984 E^4 m_{\psi }^8\nonumber\\
   &-& \cos
   (6 \theta ) (E^2-m_{\psi }^2)^6+232 E^2 m_{\psi }^{10}\nonumber\\
   &-& 2 \cos (4 \theta ) (E^2-m_{\psi
   }^2)^4 (69 E^4+6 E^2 m_{\psi }^2+50 m_{\psi }^4)\nonumber\\
   &-& \cos (2 \theta ) (E^2-m_{\psi
   }^2)^2 (975 E^8+1348 E^6 m_{\psi }^2-462 E^4 m_{\psi }^4\nonumber\\
   &+& 1932 E^2 m_{\psi }^6+1013 m_{\psi
   }^8)-1478 m_{\psi }^{12})\nonumber\\
   &+& 2 m_{\phi }^2 (E^2-m_{\psi }^2)^2 (16 E^{12} \sin ^4(\theta ) (52
   \cos (2 \theta )+\cos (4 \theta )+75)\nonumber\\
   &-& 32 E^{10} \sin ^4(\theta ) (92 \cos (2 \theta )+3 \cos (4 \theta )+193) m_{\psi }^2\nonumber\\
   &-& 4
   E^8 \sin ^2(\theta ) (1001 \cos (2 \theta )+360 \cos (4 \theta )+15 \cos (6 \theta )-6752) m_{\psi }^4\nonumber\\
   &-& 4 E^6 (-4418 \cos (2
   \theta )-414 \cos (4 \theta )+170 \cos (6 \theta )+5 \cos (8 \theta )+6961) m_{\psi }^6\nonumber\\
   &+& E^4 (-14580 \cos (2 \theta )-438 \cos (4
   \theta )+940 \cos (6 \theta )+15 \cos (8 \theta )+18671) m_{\psi }^8\nonumber\\
   &-& 2 E^2 (596 \cos (2 \theta )+2302 \cos (4 \theta )+324 \cos (6
   \theta )+3 \cos (8 \theta )-345) m_{\psi }^{10}\nonumber\\
   &+& (19438 \cos (2 \theta )+3214 \cos (4 \theta )+170 \cos (6 \theta )+\cos (8 \theta )+18649)
   m_{\psi }^{12})].
\end{eqnarray}
For brevity of notation, the argument $m$ is used as a shorthand for the masses $\{m_{\phi}, m_{\psi}\}$. This convention should be assumed whenever such characterization appears throughout this manuscript. Furthermore, in deriving the above result, we have also made use of the projectors
\begin{eqnarray}
   \sum_s \mathbf{u}_{\mu}^s(p)\bar{\mathbf{u}}_{\nu}^s(p) &=& (\slashed{p} +m_{\psi})\bigg[ \eta_{\mu \nu}-\frac{1}{3}\gamma_\mu \gamma_{\nu}-\frac{1}{3m_{\psi}}(\gamma_{\mu}p_{\nu}-\gamma_{\nu}p_{\mu})-\frac{2}{3}\frac{p_{\mu}p_{\nu}}{m_{\psi}^2} \bigg];
\end{eqnarray}
\begin{eqnarray}\label{projetvvbarprs}
   \sum_s \mathbf{v}_{\mu}^s(p)\bar{\mathbf{v}}_{\nu}^s(p) &=& (\slashed{p} - m_{\psi})\bigg[ \eta_{\mu \nu}-\frac{1}{3}\gamma_\mu \gamma_{\nu}+\frac{1}{3m_{\psi}}(\gamma_{\mu}p_{\nu}-\gamma_{\nu}p_{\mu})-\frac{2}{3}\frac{p_{\mu}p_{\nu}}{m_{\psi}^2} \bigg],
\end{eqnarray} and calculated the traces of the remaining Dirac matrices.

\begin{figure}[t]
    \centering
    \includegraphics[width=0.9\linewidth]{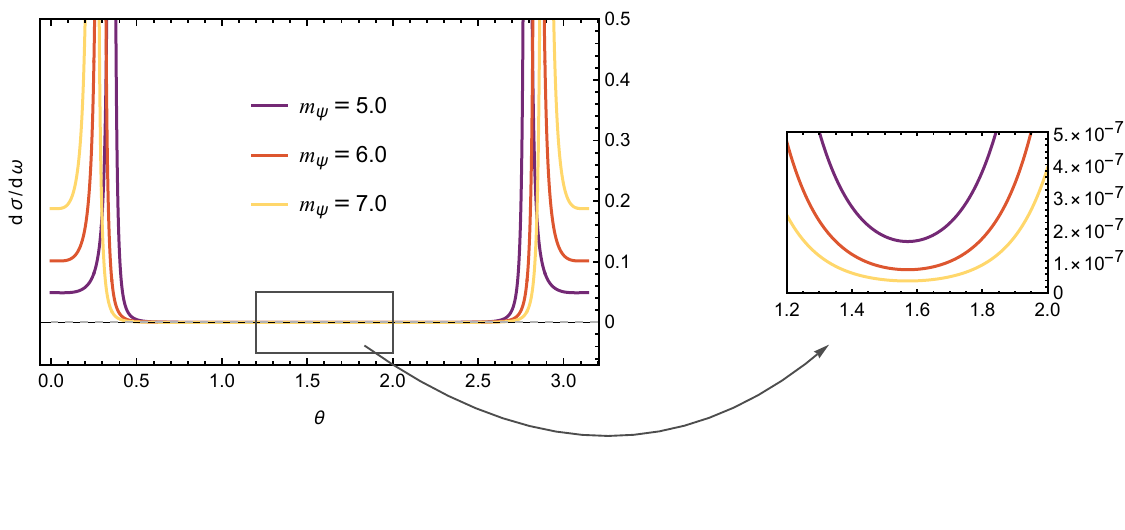}
    \caption{Angular dependence of the differential cross section at zero temperature for $E<m$ and fixed $m_{\phi}$. We have used $E=g=1.0$ and $m_{\phi}=3.0$.}
    \label{graf511}
\end{figure}

\begin{figure}[t]
    \centering
    \includegraphics[width=0.9\linewidth]{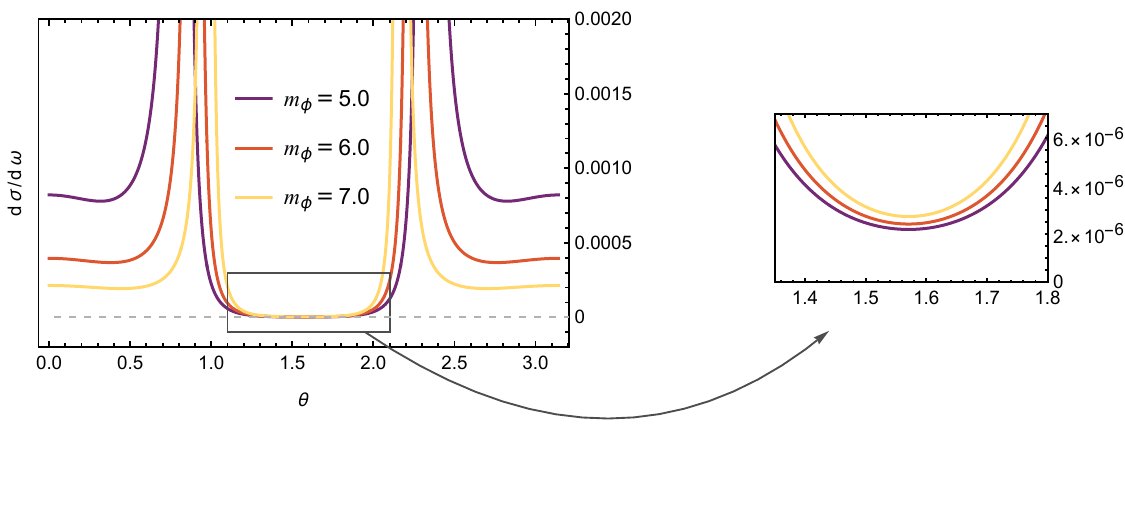}
    \caption{Angular dependence of the differential cross section at zero temperature for $E<m$ and fixed $m_{\psi}$. We have used $E=g=1.0$ and $m_{\psi}=3.0$.}
    \label{graf512}
\end{figure}

\begin{figure}[t]
    \centering
    \includegraphics[width=0.49\linewidth]{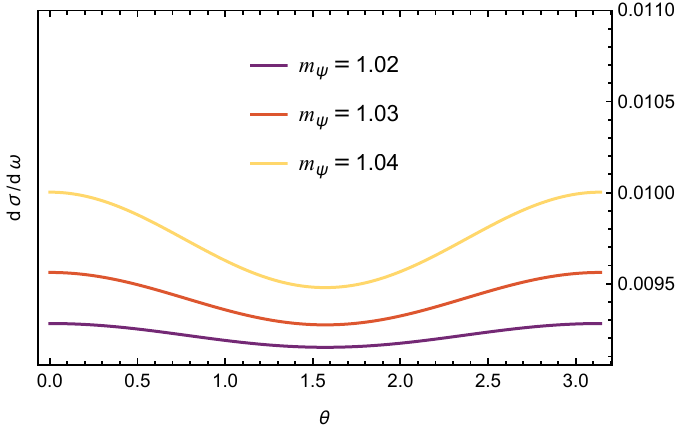}
    \includegraphics[width=0.49\linewidth]{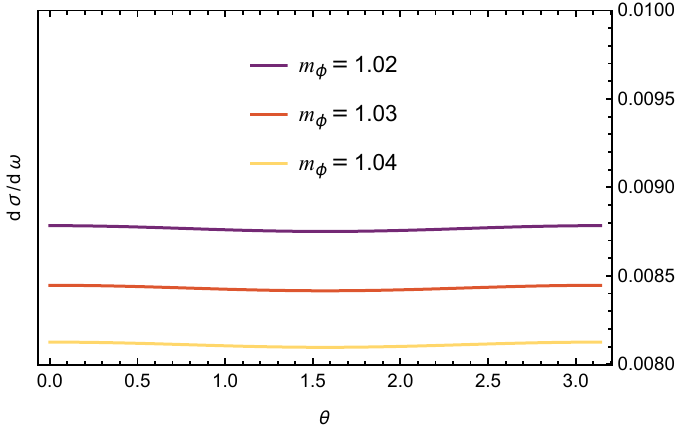}
    
    \caption{Angular dependence of the differential cross section at zero temperature for $E\approx m$, with fixed $m_{\phi}=1.01$ (left panel) and fixed $m_{\psi}=1.01$ (right panel). We have used $E=g=1.0$.}
    \label{graf513andgraf514}
\end{figure}

\begin{figure}[t]
    \centering
    \includegraphics[width=0.49\linewidth]{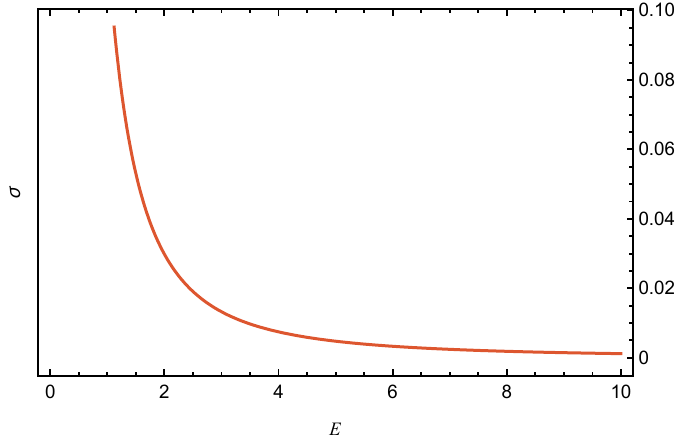}
    
    \caption{Total cross section as a function of the energy $E$ in the regime $E \approx m$. We have used $g=1.0$.}
    \label{grafeaproxm}
\end{figure}

\begin{figure}[t]
    \centering
    \includegraphics[width=0.9\linewidth]{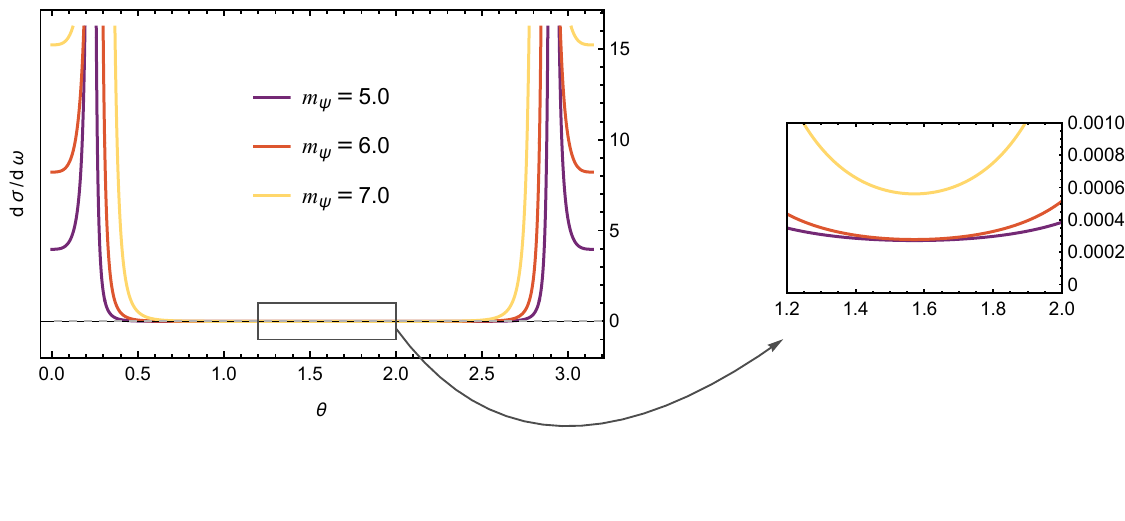}
    \caption{Angular dependence of the differential cross section at zero temperature for $E>m$ and fixed $m_{\phi}$. We have used $E=g=9.0$ and $m_{\phi}=3.0$.}
    \label{graf515}
\end{figure}

\begin{figure}[t]
    \centering
    \includegraphics[width=0.9\linewidth]{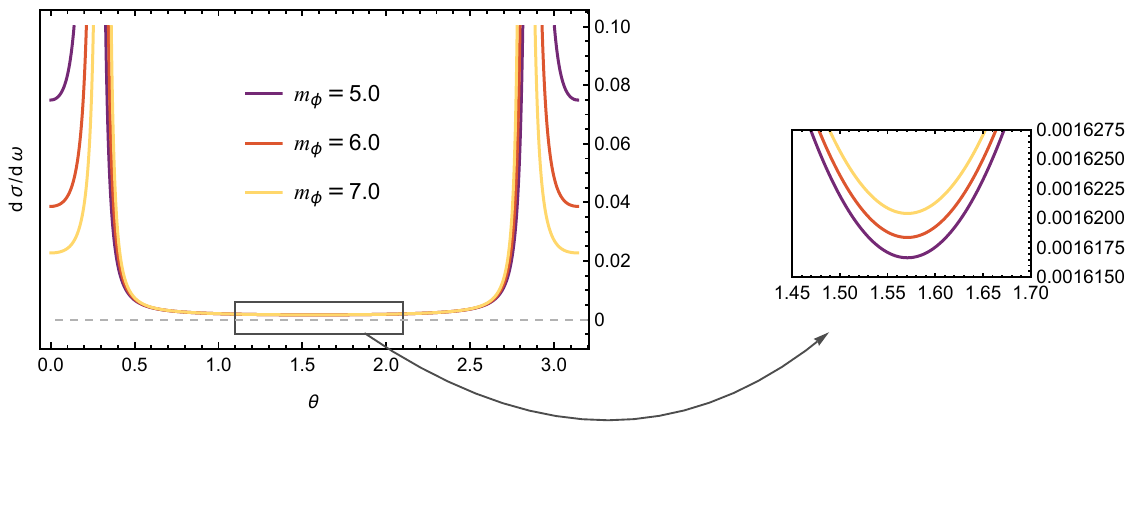}
    \caption{Angular dependence of the differential cross section at zero temperature for $E>m$ and fixed $m_{\psi}$. We have used $E=g=9.0$ and $m_{\psi}=3.0$.}
    \label{graf516}
\end{figure}

In complete analogy with the usual Yukawa theory, the result in Eq.~\eqref{secdifzerotempfinal1} is said to be evaluated in the short-range regime, where $m_{\phi} \neq 0$. In this regime, the zero-temperature differential cross section was plotted as a function of the scattering angle $\theta$ for the scenarios in which $E<m$, $E\approx m$, and $E>m$, see Figs.~\ref{graf511}-\ref{graf516}. In Fig.~\ref{graf511}, where $E<m$ and $m_{\phi}<m_{\psi}$, one observes that the larger the fermion mass, the closer the asymptotes approach the boundaries of the angular domain $0\leq\theta\leq\pi$, separating the behavior of $\mathrm{d}\sigma/\mathrm{d}\Omega$ into two distinct cases. As can be seen, in the region between each angular boundary and its nearest asymptote, the differential cross section increases with increasing $m_{\psi}$. In contrast, this behavior is reversed in the central region between the two asymptotes, since the smaller the fermion mass, the larger the differential cross section. Furthermore, the minimum of $\mathrm{d}\sigma/\mathrm{d}\Omega$ is observed to occur at $\theta \approx \pi / 2$.

In Fig.~\ref{graf512}, where $E < m$ and $m_{\phi} > m_{\psi}$, a curve profile similar to that of the previous case is observed. However, it is found that the larger the scalar boson mass, the farther the asymptotes are from the boundaries of the angular domain, which also separate the behavior of $\mathrm{d}\sigma/\mathrm{d}\Omega$ into two opposite cases. In the region between each angular boundary and its nearest asymptote, the differential cross section decreases as $m_{\phi}$ increases, whereas in the central region between the two asymptotes this behavior is reversed, i.e., the cross section tends to increase with increasing $m_{\phi}$. It is also noted that the minimum of $\mathrm{d}\sigma/\mathrm{d}\Omega$ occurs at $\theta \approx \pi / 2$.

The behavior of $\mathrm{d}\sigma/\mathrm{d}\Omega$ in the regime where $E \approx m$ is illustrated in Fig.~\ref{graf513andgraf514} for $m_{\phi} < m_{\psi}$ (left panel) and $m_{\phi} > m_{\psi}$ (right panel). Notice that, in this regime, the curves are nearly constant (horizontal lines) and lie very close to each other along the vertical direction. This suggests that the differential cross section can be approximated by
\begin{eqnarray}
   \frac{\mathrm{d} \sigma}{\mathrm{d} \Omega}  &\approx&  \frac{3\, g^4}{32\, \pi ^2\, E^2},
\end{eqnarray}
which corresponds to the result obtained in the limit $m \to E$. Integrating the expression above, we obtain a good approximation for the total cross section at zero temperature when $E \approx m$:
\begin{eqnarray}\label{sctotalzeroeaproxm111}
  \sigma \approx  \frac{3\,  g^4}{8\,  \pi \, E^2}.
\end{eqnarray}
The dependence of $\sigma$ on energy $E$ is depicted in Fig.~\ref{grafeaproxm}.

Fig.~\ref{graf515} presents the case in which $E > m$ and $m_{\phi} < m_{\psi}$. Note that although the overall shape of the curves is quite similar to those observed in Figs.~\ref{graf511} and \ref{graf512}, in this case the differential cross section increases with the fermion mass over the entire angular domain. Therefore, the inversion of behavior described previously does not occur here.

The case in which $E > m$ and $m_{\phi} > m_{\psi}$ is shown in Fig.~\ref{graf516}. As can be observed, $\mathrm{d}\sigma/\mathrm{d}\Omega$ exhibits behavior similar to that presented in Fig.~\ref{graf512}. Specifically, in the regions between each angular boundary and its nearest asymptote, the differential cross section decreases as $m_{\phi}$ increases. In contrast, in the central region between the two asymptotes, this trend is reversed, and the cross section increases with increasing $m_{\phi}$. However, in comparison with Fig.~\ref{graf512}, the asymptotes are located closer to the boundaries of the angular domain.

\begin{figure}[t]
    \centering
    \includegraphics[width=0.9\linewidth]{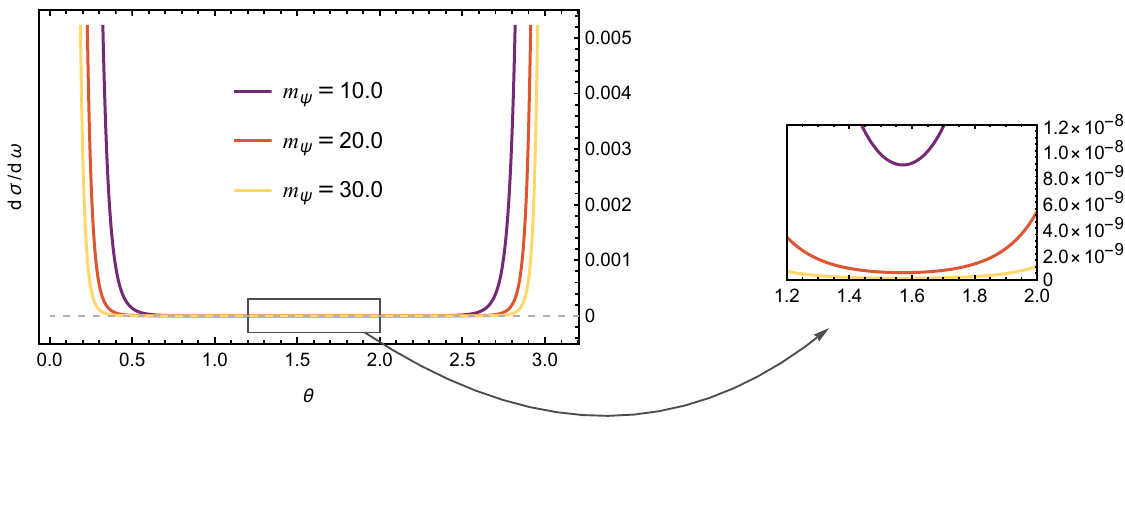}
    \caption{Angular dependence of the differential cross section at zero temperature for $E<m_{\psi}$ in the long-range regime. We have used $E=g=1.0$.}
    \label{graf521}
\end{figure}

\begin{figure}[t]
    \centering
    \includegraphics[width=0.9\linewidth]{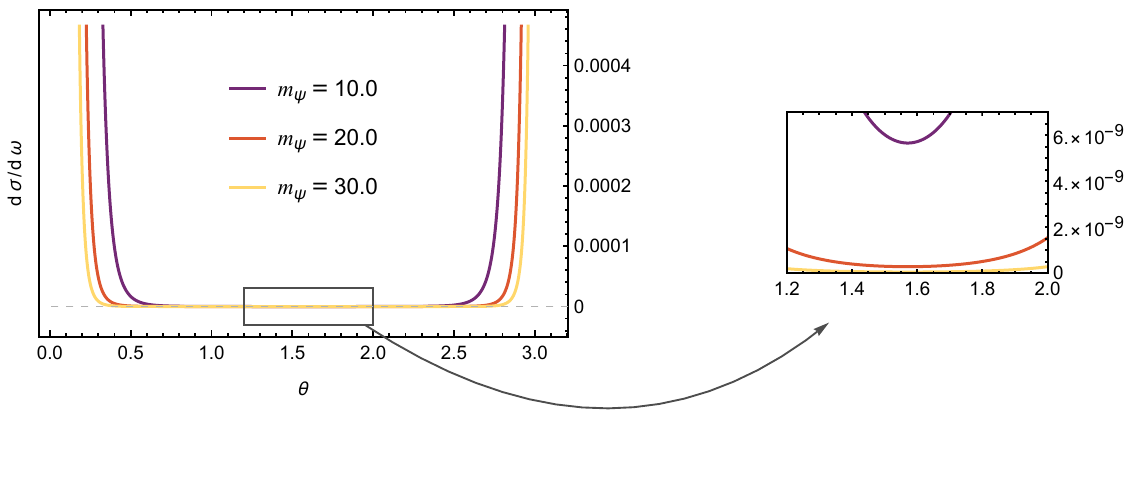}
    \caption{Angular dependence of the differential cross section at zero temperature for $E\approx m_{\psi}$ in the long-range regime. We have used $E=g=1.0$.}
    \label{graf522}
\end{figure}

\begin{figure}[t]
    \centering
    \includegraphics[width=0.7\linewidth]{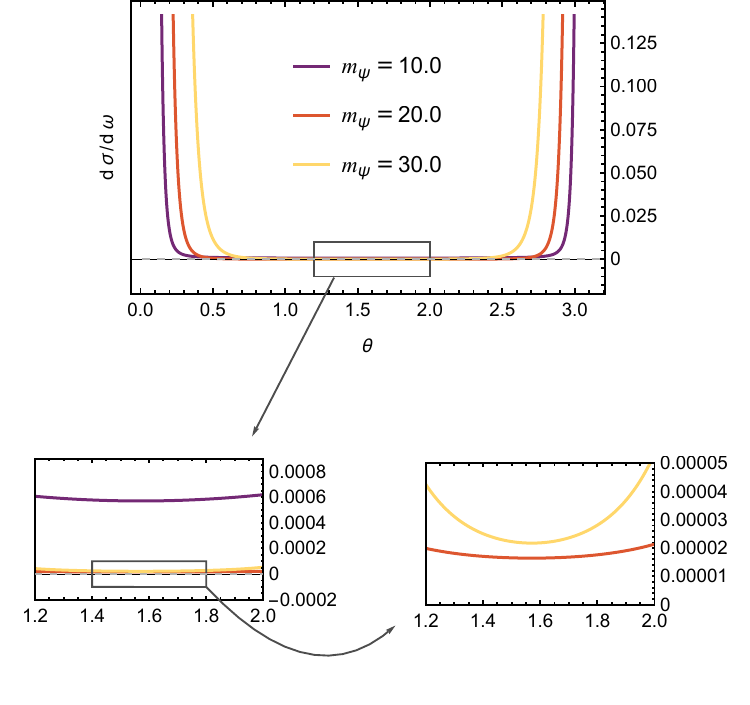}
    \caption{Angular dependence of the differential cross section at zero temperature for $E>m_{\psi}$ in the long-range regime. We have used $E=g=40.0$.}
    \label{graf523}
\end{figure}

Additionally, plots of the differential cross section at zero temperature as a function of the angle $\theta$ in the long-range regime ($m_{\phi}=0$) are shown in Figs.~\ref{graf521}-\ref{graf523}. The case $E<m_{\psi}$ is presented in Fig.~\ref{graf521}. Note that increasing the fermion mass leads to smaller values of $\mathrm{d}\sigma/\mathrm{d}\Omega$ over the entire angular range. It can also be observed that the asymptotes are shifted toward the edges of the plot, causing the differential cross section to become ill-behaved in the limits $\theta \to 0$ and $\theta \to \pi$. This latter feature is also observed in the two subsequent cases and can therefore be regarded as a consequence of the long-range regime.

Fig.~\ref{graf522} depicts the case $E \approx m_{\psi}$. In this regime, the differential cross section can be approximated by 
\begin{eqnarray}
  \frac{\mathrm{d} \sigma}{\mathrm{d} \Omega}  &\approx& \frac{g^4\,  (116 \cos (2 \theta )+3 \cos (4 \theta )+137)\,  \csc ^8(\theta )}{65536\,  \pi ^2\,  E^2\,  (E-m_{\psi})^4},
\end{eqnarray}
indicating that the closer $m_{\psi}$ is to $E$, the larger $\mathrm{d}\sigma/\mathrm{d}\Omega$ becomes. Furthermore, $\mathrm{d}\sigma/\mathrm{d}\Omega$ assumes smaller values compared to those of the previous case, although the profiles of the curves are practically the same.

Still within the long-range limit, Fig.~\ref{graf523} presents the case $E > m_{\psi}$. As can be observed, the behavior of the cross section does not remain the same throughout the entire angular domain. In fact, near the angular boundaries, $\mathrm{d}\sigma/\mathrm{d}\Omega$ increases with increasing $m_{\psi}$, whereas in the more central region an apparent lack of pattern is observed, since the yellow curve displays values larger than those of the red curve. This suggests the existence of a threshold value of $m_{\psi}$ at which the cross section, which initially decreases with increasing $m_{\psi}$, begins instead to increase with it.

\begin{figure}[t]
    \centering
    \includegraphics[width=0.49\linewidth]{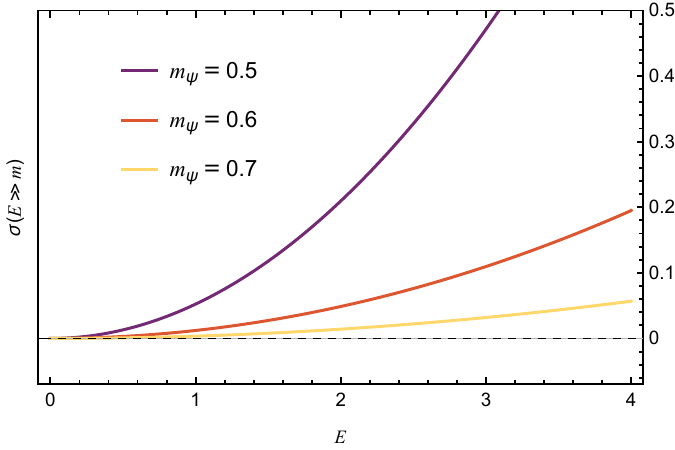}
    
    \caption{Total cross section as a function of the energy $E$ in the ultra-relativistic limit. We have used $g=1.0$.}
    \label{graf517}
\end{figure}

To conclude this section, we discuss the ultra-relativistic limit $(E\gg m)$ of Eq.~\eqref{secdifzerotempfinal1}. In this regime, the zero-temperature differential cross section is given by 
\begin{eqnarray}
  \frac{\mathrm{d} \sigma}{\mathrm{d} \Omega} &=&  \frac{g^4\, E^2\, (\cos (2 \theta )+7)}{41472\, \pi ^2\, m_{\psi}^8},
\end{eqnarray}
which can be straightforwardly integrated over the solid angle to yield the total cross section \begin{eqnarray}\label{ultarelatlimitsctotal}
    \sigma(E\gg m) = \frac{5\, g^4\, E^2  }{7776\,  \pi \, m_{\psi}^8}.
\end{eqnarray} 
As can be seen, this result does not depend explicitly on $m_{\phi}$ and is therefore identical in both the short- and long-range regimes. The dependence on the inverse of $m_{\psi}^8$ ensures that $\sigma$ always decreases as the fermion mass increases. Furthermore, with respect to its energy dependence, Eq.~\eqref{ultarelatlimitsctotal} exhibits behavior opposite to that found in the regime where $E \approx m$ (see Eq.~\eqref{sctotalzeroeaproxm111}), as it is directly proportional to $E^2$. Fig.~\ref{graf517} presents a plot of $\sigma$ as a function of $E$ for selected values of $m_{\psi}$ in the ultra-relativistic limit.

\section{Analysis at finite temperature} \label{TFD}

\subsection{The thermofield dynamics formalism}

The TFD formalism is a real-time framework that enables the investigation of temperature effects in systems described by quantum field theory. Its construction is based on a duplication of the degrees of freedom of the system, where for each operator $\mathcal{O}$ in a space $\mathcal{S}$, there exists an operator $\widehat{\mathcal{O}}$ in a space $\widehat{\mathcal{S}}$, defined by

\begin{eqnarray}\label{ohatomenosotilde}
\widehat{\mathcal{O}} \equiv \mathcal{O} - \widetilde{\mathcal{O}}.
\end{eqnarray}
Here, $\widetilde{\mathcal{O}}$ is an operator in a space $\widetilde{\mathcal{S}}$, conceived as a copy of the original space $\mathcal{S}$  where physical observables are indeed defined. $\widehat{\mathcal{S}}$, in turn, characterizes the duplicated space in which the TFD formalism operates. The tilde and non-tilde operators, although independent in the sense that they operate solely within their respective domains, are in fact connected through the conjugation rules

\begin{eqnarray}\label{rct1}
\widetilde{(\mathcal{O}_i\mathcal{O}_j)} &=& \widetilde{\mathcal{O}}_i\widetilde{\mathcal{O}}_j ;\\
\widetilde{(a\mathcal{O}_i+b\mathcal{O}_j)} &=& a^{\ast}\widetilde{\mathcal{O}}_i+b^{\ast}\widetilde{\mathcal{O}}_j;\\
(\widetilde{\mathcal{O}}_i)^{\dagger} &=& \widetilde{(\mathcal{O}_i^{\dagger})};\\
\label{rct4}
\widetilde{(\widetilde{\mathcal{O}}_i)} &=& \xi \mathcal{O}_i,
\end{eqnarray}
where $\xi=1\, (-1)$ for bosons (fermions). Within this doubled structure, the TDF formalism allows for the construction of thermal operators in terms of the non-thermal ones through what we refer to as Bogoliubov transformations. In general, we write
\begin{eqnarray}\label{opefermionbetazero1}
  a(\mathbf{k},\beta) &=& u_j(\mathbf{k},\beta)a(\mathbf{k}) - v_j(\mathbf{k},\beta)\widetilde{a}^{\dagger}(\mathbf{k}); \\ \label{opefermionbetazero2}
a^{\dagger}(\mathbf{k},\beta) &=& u_j(\mathbf{k},\beta)a^{\dagger}(\mathbf{k}) - v_j(\mathbf{k},\beta)\widetilde{a}(\mathbf{k}); \\ \label{opefermionbetazero3}
\widetilde{a}(\mathbf{k},\beta) &=& u_j(\mathbf{k},\beta)\widetilde{a}(\mathbf{k}) -\xi\,  v_j(\mathbf{k},\beta)a^{\dagger}(\mathbf{k});\\ \label{opefermionbetazero4}
\widetilde{a}^{\dagger}(\mathbf{k},\beta) &=& u_j(\mathbf{k},\beta)\widetilde{a}^{\dagger}(\mathbf{k}) - \xi\,  v_j(\mathbf{k},\beta)a(\mathbf{k}),
 \end{eqnarray}
 where
\begin{eqnarray}\label{defufermioncostheta1}
  u_j(\mathbf{k},\beta) &=& e^{\beta\, |k_0|/2}\, v_j(\mathbf{k},\beta), 
 \end{eqnarray}
 and
\begin{eqnarray}
  \label{defufermioncostheta2}
  v_j(\mathbf{k},\beta) &=& \frac{1}{\sqrt{e^{\beta\, |k_0|}-\xi}},
\end{eqnarray} 
 with $j = B$ for bosons and $j = F$ for fermions. In the above expressions, $\beta=1/T$ ($T$ being the temperature), $k_0$ is the zero component associated with the mode of four-momentum $k$, and $a\, (\widetilde{a})$ and $a^{\dagger}\, (\widetilde{a}^{\dagger})$ stand for annihilation and creation operators in the space $\mathcal{S}$ ($\widetilde{\mathcal{S}}$), respectively. It can also be shown that the algebra satisfied by such operators is similar to the conventional one, that is, for fermions,
 \begin{eqnarray}
\label{algfermion1}
   \left\{ a^{\lambda}(\mathbf{k},\beta),a^{\lambda'\, \dagger}(\mathbf{k'},\beta)\right\} &=& (2\pi)^3\delta^3(\mathbf{k}-\mathbf{k'})\delta^{\lambda \lambda'};\\
\label{algfermion2} \left\{ \widetilde{a}^{\, \lambda}(\mathbf{k},\beta),\widetilde{a}^{\, \lambda'\, \dagger}(\mathbf{k'},\beta)\right\} &=& (2\pi)^3\delta^3(\mathbf{k}-\mathbf{k'})\delta^{\lambda \lambda'},
 \end{eqnarray}
  while for bosons, the brackets representing the anticommutation operation must be replaced by the commutation brackets. It is important to mention that all other commutation or anticommutation relations are null, and that the superscripts $\lambda$ and $\lambda'$ have been implemented to account for some physically relevant property of interest, such as the fermion spin or the photon polarization state, for example.

In the TFD formalism, the thermal differential cross section is directly proportional to the square of the temperature-dependent scattering amplitude $\widehat{\mathcal{M}}(\beta)$, where
\begin{eqnarray}\label{mhatmbetamenosmtilbetadef}
 \widehat{\mathcal{M}}(\beta) = \mathcal{M}(\beta) - \widetilde{\mathcal{M}}(\beta) 
\end{eqnarray}
with
\begin{equation}\label{matrizmbetaoriginal}
   \mathcal{M}(\beta) \equiv \sum_{n=0}^{\infty}\frac{(-i)^n}{n!}\int\, d^4z_1\cdots d^4z_n\, {}_{\beta}\langle \, f\, |\mathcal{T}\left[ {\mathcal{H}}_{int}(z_1)\cdots {\mathcal{H}}_{int}(z_n) \right]|\, i\, \rangle_{\beta}
\end{equation} 
and
\begin{equation}\label{matrizbetatil}
   \widetilde{\mathcal{M}}(\beta) \equiv \sum_{n=0}^{\infty}\frac{\widetilde{(-i)^n}}{n!}\int\, d^4z_1\cdots d^4z_n\, {}_{\beta}\langle \, f\, |\mathcal{T}\left[ \widetilde{\mathcal{H}}_{int}(z_1)\cdots \widetilde{\mathcal{H}}_{int}(z_n) \right]|\, i\, \rangle_{\beta},
\end{equation}
in full agreement with Eq. \eqref{ohatomenosotilde}. Here, ${\mathcal{H}}_{int}$ ($\widetilde{\mathcal{H}}_{int}$) is the interaction Hamiltonian density in the space ${\mathcal{S}}$ ($\widetilde{\mathcal{S}}$), and $\mathcal{T}$ is the time-ordering operator. For the scattering process of interest, the initial and final particle states in Eqs.~\eqref{matrizmbetaoriginal} and \eqref{matrizbetatil} are respectively defined by 
\begin{eqnarray}
  |\, i\, \rangle_{\beta} &\equiv& \sqrt{2E(\mathbf{p}_1)\, 2E(\mathbf{p}_2)}\, a^{s_1\, \dagger}(\mathbf{p}_1,\beta)\, a^{s_2 \, \dagger}(\mathbf{p}_2,\beta)\, |\, 0(\beta)\, \rangle
\end{eqnarray}
and 
\begin{eqnarray}
    {}_{\beta}\langle \, f\, | &\equiv& \langle \, 0(\beta)\, | \, a^{r_2}(\mathbf{k}_2,\beta)\, a^{r_1}(\mathbf{k}_1,\beta) \sqrt{2E(\mathbf{k}_1)\,  2E(\mathbf{k}_2)},
\end{eqnarray}
where $|\, 0(\beta)\, \rangle$ is the thermal vacuum state. It should be emphasized here that both ${\mathcal{H}}_{int}$ and $\widetilde{\mathcal{H}}_{int}$ are expressed in terms of products of fields whose plane-wave expansions, conventional in their respective spaces, do not depend explicitly on thermal operators. Therefore, in order to properly treat Eqs.~\eqref{matrizmbetaoriginal} and \eqref{matrizbetatil}, it is necessary to apply the Bogoliubov transformations given in Eqs.~\eqref{opefermionbetazero1}-\eqref{opefermionbetazero4}. 

In practice, the procedure for constructing the matrix $\widehat{\mathcal{M}}$ is closely analogous to that discussed in the previous section. Once the Feynman diagrams for the process of interest have been drawn - which are common to both spaces, namely $\mathcal{S}$ and $\widetilde{\mathcal{S}}$ - we must apply the corresponding Feynman rules in order to write the matrices $\mathcal{M}$ and $\widetilde{\mathcal{M}}$ in an appropriate form. In our case, the diagrams to be considered are the same as those shown in Fig.~\ref{diagrelevantes}, while the Feynman rules are modified as a direct consequence of the definitions given in Eqs.~\eqref{matrizmbetaoriginal} and~\eqref{matrizbetatil}. In general, two modifications are observed. The first one occurs in the scalar propagator, which in the spaces $\mathcal{S}$ and $\widetilde{\mathcal{S}}$ is defined by
\begin{eqnarray}
   D(q,\beta)=  \langle\, 0(\beta)\, |\mathcal{T}\left[ \phi(x)\phi(y) \right]|\, 0(\beta)\, \rangle ,
\end{eqnarray} 
and
\begin{eqnarray}
   \widetilde{D}(q,\beta)=  \langle\, 0(\beta)\, |\mathcal{T}\left[ \widetilde{\phi}(x)\widetilde{\phi}(y) \right]|\, 0(\beta)\, \rangle ,
\end{eqnarray}
respectively, with 
\begin{eqnarray}
  \phi(x) = \int \frac{\mathrm{d}^3k}{(2\pi)^3}\, \frac{1}{\sqrt{2E(\mathbf{k})}}\left[ a(\mathbf{k})e^{-ik\cdot x} + a^{\dagger}(\mathbf{k})e^{ik\cdot x} \right],
\end{eqnarray}
representing the usual plane-wave expansion for the scalar field. Explicitly, we may write
\begin{eqnarray}\label{propqbetaterms}
  D(q,\beta) = \frac{i}{q^2-m_{\phi}^2} + 2\pi\, v_B^2(\mathbf{q},\beta)\delta(q^2-m_{\phi}^2)
\end{eqnarray}
and
\begin{eqnarray}\label{propqbetatermstil}
    \widetilde{D}(q,\beta) = \frac{-i}{q^2-m_{\phi}^2} + 2\pi\, v_B^2(\mathbf{q},\beta)\delta(q^2-m_{\phi}^2)
\end{eqnarray}
where we have made use of the conjugation rules given in Eqs.~\eqref{rct1}-\eqref{rct4} and the Bogoliubov transformations in Eqs.~\eqref{opefermionbetazero1}-\eqref{opefermionbetazero4}. The presence of the Dirac delta function in the second term of the thermal propagators is a distinctive feature of the TFD formalism. It gives rise to a type of apparent singularity, which can be properly circumvented by means of a regularization scheme for the delta function and its derivatives, explicitly given by
\begin{eqnarray}
  \label{propriedadedaderivadadadeltaprop}
 2\pi i \frac{1}{n!}\frac{\partial^n}{\partial x^n }\delta (x) = \left( -\frac{1}{x+i\epsilon} \right)^{n+1} - \left( -\frac{1}{x-i\epsilon} \right)^{n+1}.
\end{eqnarray} In fact, this has been effectively employed in the derivation of Eqs.~\eqref{propqbetaterms} and~\eqref{propqbetatermstil}. 

The second modification in the Feynman rules appears in the expressions for the external legs of the diagrams. For instance, an external spin-$3/2$ fermion line at the initial state of the process is represented in the spaces $\mathcal{S}$ and $\widetilde{\mathcal{S}}$ by the expressions $u_F(\mathbf{p},\beta)\mathbf{u}_{\mu}^s(p)$ and $-v_F(\mathbf{p},\beta)\bar{\mathbf{u}}_{\mu}^{\ast \, s}(p)$, respectively. Note that this result once again follows directly from the field expansion in Eq.~\eqref{campopsimuxexpan} after the application of Eqs.~\eqref{rct1}-\eqref{opefermionbetazero4}. 

Finally, the interaction vertex in the original space does not undergo any modification when compared to that of Eq.~\eqref{verticemunuetagammagamma}. Nevertheless, we must also consider the vertex in the tilde space, which is simply given by $\widetilde{V}^{\mu \nu}=(V^{\mu \nu})^{\ast}$. 

\subsection{Yukawa-like scattering revisited}

Within the TFD formalism, the thermal differential cross section, when once again evaluated in the center-of-mass frame, is defined as \begin{eqnarray}\label{secdifbetacmsomaspinmhatcmbetaquadrado}
      \left( \frac{\mathrm{d} \sigma}{\mathrm{d} \Omega}   \right)_{\beta} = \frac{1}{64\, \pi^2 \, E_{cm}^2}\, \frac{1}{16}\sum_{spin}|\widehat{\mathcal{M}}(\beta)|^2.
\end{eqnarray}
Note that, similarly to the zero-temperature case, since both diagrams in Fig.~\ref{diagrelevantes} must be taken into account, one should write
\begin{eqnarray}
    \widehat{\mathcal{M}}(\beta) = \widehat{\mathcal{M}}_t(\beta)+\widehat{\mathcal{M}}_u(\beta),
\end{eqnarray}
where, according to Eq.~\eqref{ohatomenosotilde}, 
\begin{eqnarray}
    \widehat{\mathcal{M}}_t(\beta) = \mathcal{M}_t(\beta) - \widetilde{\mathcal{M}}_t(\beta); 
\end{eqnarray}
\begin{eqnarray}
    \widehat{\mathcal{M}}_u(\beta) = \mathcal{M}_u(\beta) - \widetilde{\mathcal{M}}_u(\beta), 
\end{eqnarray}
with
\begin{eqnarray}
    \mathcal{M}_t(\beta) &=& \frac{1}{q^2-m_{\phi}^2} \left(\prod v_F\right)\left(\prod \frac{u_F}{v_F}\right)\left[ 1-2 \pi i\,  v_B^2(\mathbf{q},\beta)\, (q^2-m_{\phi}^2)\, \delta(q^2-m_{\phi}^2)\right]\nonumber\\
    &\times& \mathbf{\bar{u}}_{\mu}^{r_1}(k_1)\, V^{\mu \nu}\, \mathbf{u}_{\nu}^{s_1}(p_1)\, \mathbf{\bar{u}}_{\mu'}^{r_2}(k_2)\, V^{\mu' \nu'}\, \mathbf{u}_{\nu'}^{s_2}(p_2);
\end{eqnarray}
\begin{eqnarray}
    \widetilde{\mathcal{M}}_t(\beta) &=& \frac{1}{q^2-m_{\phi}^2} \left(\prod v_F\right)\left[ 1+2 \pi i\,  v_B^2(\mathbf{q},\beta)\, (q^2-m_{\phi}^2)\, \delta(q^2-m_{\phi}^2)\right]\nonumber\\
    &\times& \mathbf{\bar{u}}_{\mu}^{r_1}(k_1)\, V^{\mu \nu}\, \mathbf{u}_{\nu}^{s_1}(p_1)\, \mathbf{\bar{u}}_{\mu'}^{r_2}(k_2)\, V^{\mu' \nu'}\, \mathbf{u}_{\nu'}^{s_2}(p_2);
\end{eqnarray}
\begin{eqnarray}
    \mathcal{M}_u(\beta) &=& \frac{-1}{q'^2-m_{\phi}^2} \left(\prod v_F\right)\left(\prod \frac{u_F}{v_F}\right)\left[ 1-2 \pi i\,  v_B^2(\mathbf{q}',\beta)\, (q'^2-m_{\phi}^2)\, \delta(q'^2-m_{\phi}^2)\right]\nonumber\\
    &\times& \mathbf{\bar{u}}_{\mu}^{r_2}(k_2)\, V^{\mu \nu}\, \mathbf{u}_{\nu}^{s_1}(p_1)\, \mathbf{\bar{u}}_{\mu'}^{r_1}(k_1)\, V^{\mu' \nu'}\, \mathbf{u}_{\nu'}^{s_2}(p_2);
\end{eqnarray}
\begin{eqnarray}
    \widetilde{\mathcal{M}}_u(\beta) &=& \frac{-1}{q'^2-m_{\phi}^2} \left(\prod v_F\right)\left[ 1+2 \pi i\,  v_B^2(\mathbf{q}',\beta)\, (q'^2-m_{\phi}^2)\, \delta(q'^2-m_{\phi}^2)\right]\nonumber\\
    &\times& \mathbf{\bar{u}}_{\mu}^{r_2}(k_2)\, V^{\mu \nu}\, \mathbf{u}_{\nu}^{s_1}(p_1)\, \mathbf{\bar{u}}_{\mu'}^{r_1}(k_1)\, V^{\mu' \nu'}\, \mathbf{u}_{\nu'}^{s_2}(p_2).
\end{eqnarray}
The last four expressions were written according to the modified Feynman rules discussed in the previous subsection, and we further define
\begin{eqnarray}
    \prod v_F \equiv v_F(\mathbf{p_1},\beta)\, v_F(\mathbf{p_2},\beta)\, v_F(\mathbf{k_1},\beta)\, v_F(\mathbf{k_2},\beta), 
\end{eqnarray} 
and
\begin{eqnarray}
    \prod \frac{u_F}{v_F} \equiv \frac{u_F(\mathbf{p_1},\beta)\, u_F(\mathbf{p_2},\beta)\, u_F(\mathbf{k_1},\beta)\, u_F(\mathbf{k_2},\beta)}{v_F(\mathbf{p_1},\beta)\, v_F(\mathbf{p_2},\beta)\, v_F(\mathbf{k_1},\beta)\, v_F(\mathbf{k_2},\beta)}.
\end{eqnarray}
Therefore, bearing in mind Eqs.~\eqref{setcentermassframe33}-\eqref{projetvvbarprs}, we can write the temperature-dependent differential cross section for the scattering process described in Eq.~\eqref{fermionfermionsctteringyukawa} within the framework of the Rarita-Schwinger theory with Yukawa-like coupling as
\begin{eqnarray}\label{scdiftermalfinal12321}
    \left( \frac{\mathrm{d} \sigma}{\mathrm{d} \Omega} \right)_{\beta} &=& \frac{g^4}{41472\, \pi ^2\, E^2\, m^8_{\psi}}\big\{    \big[\, \Pi_1(m,E,\theta)\, \delta \left(\zeta_{-}\right)^2+\Pi_2(m,E,\theta)\, \delta \left(\zeta_{+}\right)^2 \nonumber\\
    &+& \Pi_3(m,E,\theta)\, \delta \left(\zeta_{+}\right)\, \delta \left(\zeta_{-}\right)\, \big]\,  \Gamma(E,\beta) + \Pi_0(m,E,\theta)\, \Theta(E,\beta) \big\}, 
\end{eqnarray}
with 
\begin{eqnarray}
    \Pi_1(m,E,\theta) &=& 128\, \pi ^2\, [8 E^6 \sin ^6\textstyle{\left(\frac{\theta }{2}\right)}+4 E^4 \sin ^4\textstyle{\left(\frac{\theta }{2}\right)} (3 \cos (\theta )+2) m_{\psi
   }^2\nonumber\\
   &+& E^2 \sin ^2\textstyle{\left(\frac{\theta }{2}\right)} (8 \cos (\theta )+3 \cos (2 \theta )+10) m_{\psi }^4\nonumber\\
   &+&\cos ^2\textstyle{\left(\frac{\theta }{2}\right)} (2 \cos (\theta )+\cos (2 \theta )+6) m_{\psi }^6]^2;
\end{eqnarray}

\begin{eqnarray}
    \Pi_2(m,E,\theta) &=& 128\, \pi ^2\, [8 E^6 \cos ^6\textstyle{\left(\frac{\theta }{2}\right)}+4 E^4 \cos ^4\textstyle{\left(\frac{\theta }{2}\right)} (2-3 \cos (\theta ))  m_{\psi
   }^2\nonumber\\
   &+&E^2 \cos ^2\textstyle{\left(\frac{\theta }{2}\right)} (-8 \cos (\theta )+3 \cos (2 \theta )+10)  m_{\psi }^4\nonumber\\
   &+&\sin ^2\textstyle{\left(\frac{\theta
   }{2}\right)} (-2 \cos (\theta )+\cos (2 \theta )+6) m_{\psi }^6]^2;
\end{eqnarray}

\begin{eqnarray}
    \Pi_3(m,E,\theta) &=& 4\, \pi ^2\, [ 32 E^{12} \sin ^6(\theta )+32 E^{10} \sin ^4(\theta ) (3 \cos (2 \theta )-31) m_{\psi }^2\nonumber\\
    &+& 4 E^8 \sin ^2(\theta ) (-502
   \cos (2 \theta )+15 \cos (4 \theta )+1159) m_{\psi }^4\nonumber\\
   &+& 4 E^6 (1353 \cos (2 \theta )-192 \cos (4 \theta )+5 \cos (6 \theta )-1454)
   m_{\psi }^6\nonumber\\
   &+& E^4 (-2971 \cos (2 \theta )+502 \cos (4 \theta )-15 \cos (6 \theta )+3060) m_{\psi }^8\nonumber\\
   &+& 2 E^2 (227 \cos (2
   \theta )-62 \cos (4 \theta )+3 \cos (6 \theta )-528) m_{\psi }^{10}\nonumber\\
   &-& (81 \cos (2 \theta )+\cos (6 \theta )-82) m_{\psi }^{12} ];
\end{eqnarray}

\begin{eqnarray}
    \Gamma(E,\beta) &=& \frac{\cosh ^2(\beta  E) \text{csch}^2\left(\frac{\beta  E}{2}\right) (\sinh (\beta  E)+\cosh (\beta 
   E))}{(\sinh (\beta  E)+\cosh (\beta  E)+1)^6};
\end{eqnarray}

\begin{eqnarray}
    \Theta(E,\beta) &=& \tanh ^2\left(\frac{\beta  E}{2}\right);
\end{eqnarray}

\begin{eqnarray}
    \zeta_{\pm}(m,E,\theta) = m_{\phi }^2-4 (\cos (\theta )\pm 1)^2 (E^2-m_{\psi }^2)^2, 
\end{eqnarray} and $\Pi_0(m,E,\theta)$ defined in Eq.~\eqref{deffuncpizerometheta}. In obtaining the above result, all involved Dirac matrix traces have been evaluated.

\begin{figure}[t]
    \centering
    \includegraphics[width=0.49\linewidth]{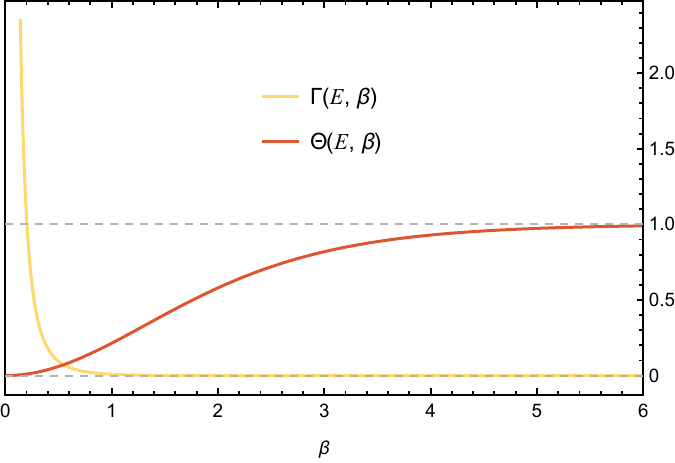}
    
    \caption{Dependence of the functions $\Gamma$ and $\Theta$ on $\beta$. We have used $E=1.0$.}
    \label{graftempfunc}
\end{figure}

Fig.~\ref{graftempfunc} displays the $\beta$-dependence of the functions $\Gamma$ and $\Theta$. As can be observed, at very high temperatures (i.e., very small $\beta$), thermal corrections to the differential cross section become extremely significant and therefore cannot be neglected. In this regime, it can be shown that $\left( \mathrm{d} \sigma / \mathrm{d} \Omega \right)_{\beta}$ is directly proportional to $T^2$. Furthermore, note that $\Gamma$ rapidly decreases to zero as the temperature is lowered, indicating that $\left( \mathrm{d} \sigma / \mathrm{d} \Omega \right)_{\beta}$ for small values of $T$ can be approximated by
\begin{eqnarray}
    \left( \frac{\mathrm{d} \sigma}{\mathrm{d} \Omega} \right)_{\beta} &\approx& \Theta(E,\beta)\,  \frac{\mathrm{d} \sigma}{\mathrm{d} \Omega}, 
\end{eqnarray} 
where $ \mathrm{d} \sigma / \mathrm{d} \Omega$ is the zero-temperature differential cross
section, as given in Eq.~\eqref{secdifzerotempfinal1}. Additionally, since $\Gamma \to 0$ and $\Theta \to 1$ as $T \to 0$, the expression in Eq.~\eqref{scdiftermalfinal12321} naturally reduces to its zero-temperature counterpart, Eq.~\eqref{secdifzerotempfinal1}.

\section{Conclusion}\label{conclusion}

We have investigated fermion-fermion scattering in a Rarita-Schwinger model, incorporating a Yukawa-like interaction via the replacement $m \rightarrow m_{\psi} + g\phi$ in the free massive spin-$3/2$ Lagrangian. The analysis was carried out at both zero and finite temperatures, with the latter implemented through the Thermofield Dynamics (TFD) formalism. The scattering processes were computed at tree level, considering the center-of-mass frame for both temperature regimes. The results obtained are as general as possible, although different scenarios regarding particle masses and energies were also examined. In general, the differential cross section, whether at zero or finite temperature, can be divided into two regimes: the short-range regime ($m_{\phi} \neq 0$) and the long-range regime ($m_{\phi} = 0$).

In the short-range regime at $T=0$, the case $E < m$ ($m$ denoting either $m_{\phi}$ or $m_{\psi}$) with $m_{\phi} < m_{\psi}$ and fixed $m_{\phi}$, for instance, shows that the values of the scattering angle $\theta$ for which $\mathrm{d}\sigma/\mathrm{d}\Omega$ is ill-behaved depend strongly on the fermion mass. Indeed, the corresponding asymptotes tend to shift toward the boundaries of the angular domain $0 \leq \theta \leq \pi$ as $m_{\psi}$ increases. Moreover, these asymptotes separate the behavior of the differential cross section into two distinct regions, depending on whether $\mathrm{d}\sigma/\mathrm{d}\Omega$ increases or decreases with increasing $m_{\psi}$. Specifically, in the central region between the two asymptotes, $\mathrm{d}\sigma/\mathrm{d}\Omega$ decreases as $m_{\psi}$ increases, whereas near the edges of the angular domain it increases with increasing $m_{\psi}$. Our results also indicate that the angular values for which the differential cross section is ill-behaved depend sensitively on the scalar boson mass. When $E < m$ and $m_{\phi} > m_{\psi}$, with $m_{\psi}$ fixed in the short-range regime at zero temperature, the asymptotes move toward one another, approaching the central region of the angular domain as $m_{\phi}$ increases. In contrast to the previous case, although the overall shape of the curves remains similar, $\mathrm{d}\sigma/\mathrm{d}\Omega$ exhibits the opposite behavior: it decreases near the edges of the angular domain and increases in the central region between the asymptotes as $m_{\phi}$ grows.

The case $E \approx m$ was also investigated. In this regime, we show that $\mathrm{d}\sigma/\mathrm{d}\Omega$ displays no dependence on the masses $m_{\phi}$ and $m_{\psi}$ or on the scattering angle $\theta$, remaining essentially constant throughout the entire angular domain. In this limit, we derived an approximate expression for the total cross section, revealing a dependence proportional to $E^{-2}$. It is important to emphasize that the opposite angular behaviors of $\mathrm{d}\sigma/\mathrm{d}\Omega$ observed in the two first analyses discussed above are not universal, but rather case-dependent. This becomes evident in the case $E > m$ with $m_{\phi} < m_{\psi}$ and fixed $m_{\psi}$ in the short-range regime at zero temperature, where the differential cross section assumes larger values for larger $m_{\psi}$ over the entire angular domain. In the long-range regime, in turn, the case $E < m_{\psi}$ shows that $\mathrm{d}\sigma/\mathrm{d}\Omega$ decreases across the entire angular domain as $m_{\psi}$ increases. Our results also suggest what appears to be a general feature of the long-range regime, namely that the differential cross section is always ill-behaved at the boundaries of the angular domain. 

The behavior of the differential cross section in the ultrarelativistic limit was likewise analyzed. In this regime, no dependence on the scalar boson mass is observed, and thus there is no distinction between the short- and long-range regimes. We also obtained an expression for the total cross section, showing that it scales with the square of the energy and with the inverse eighth power of the fermion mass.

From our finite-temperature ($T \neq 0$) result, we conclude that thermal corrections to the differential cross section become highly significant at very
high temperatures, and therefore cannot be neglected. In this regime, we demonstrate that $\left( \mathrm{d} \sigma / \mathrm{d} \Omega \right)_{\beta}$ scales with the square of the temperature. Finally, for sufficiently small $T$, the approximate finite-temperature result can be written in terms of its zero-temperature counterpart through a multiplicative, $T$-dependent factor that approaches unity in the limit $T \to 0$. In this way, the zero-temperature result is naturally recovered.

\section*{Acknowledgments}
\hspace{0.5cm} MCA would like to thank FUNCAP (Process No. DC3-0235-00076.01.00/24) and CNPq (Process No. 304145/2025-4) for financial support. JGL would like to thank CAPES (Finance Code 001). JF would like to thank FUNCAP (Grant PRONEM PNE0112-00085.01.00/16) and CNPq (Grant 304485/2023-3) for financial support. TM would like to thank FAPEAL (Project No. E:60030.0000002341/2022) and CNPq (Project No. 309360/2025-0) for financial support.

\vspace{0.5cm}

\center{\bf No Data associated in the manuscript}

\appendix

\global\long\def\link#1#2{\href{http://eudml.org/#1}{#2}}
 \global\long\def\doi#1#2{\href{http://dx.doi.org/#1}{#2}}
 \global\long\def\arXiv#1#2{\href{http://arxiv.org/abs/#1}{arXiv:#1 [#2]}}
 \global\long\def\arXivOld#1{\href{http://arxiv.org/abs/#1}{arXiv:#1}}


\end{document}